\documentclass{Interspeech}

\usepackage[capitalize]{cleveref}

\usepackage{tikz}
\usepackage{arydshln}

\interspeechcameraready

\title{Cross-Modal Watermarking for Authentic Audio Recovery and Tamper Localization in Synthesized Audiovisual Forgeries}

\author[affiliation={1}]{Minyoung}{Kim}
\author[affiliation={1}]{Sehwan}{Park}
\author[affiliation={2}]{Sungmin}{Cha}
\author[affiliation={1}]{Paul Hongsuck}{Seo}

\affiliation{Dept. of CSE}{Korea University}{Republic of Korea}
\affiliation{}{New York University}{USA}
\email{\{omniverse186, shp216, phseo\}@korea.ac.kr \\ sungmin.cha@nyu.edu} 


\usepackage{comment}
\usepackage{pifont}

\begin{document}

\maketitle


\begin{abstract}
    Recent advances in voice cloning and lip synchronization models have enabled Synthesized Audiovisual Forgeries (SAVFs), where both audio and visuals are manipulated to mimic a target speaker. This significantly increases the risk of misinformation by making fake content seem real. To address this issue, existing methods detect or localize manipulations but cannot recover the authentic audio that conveys the semantic content of the message. This limitation reduces their effectiveness in combating audiovisual misinformation. In this work, we introduce the task of Authentic Audio Recovery (AAR) and Tamper Localization in Audio (TLA) from SAVFs and propose a cross-modal watermarking framework to embed authentic audio into visuals before manipulation. This enables AAR, TLA, and a robust defense against misinformation. Extensive experiments demonstrate the strong performance of our method in AAR and TLA against various manipulations, including voice cloning and lip synchronization.\footnote{The code is available at: \url{https://eurominyoung186.github.io/CMW_SAVF/}}
\end{abstract}
\section{Introduction}
\label{sec:intro}

Recent advancements in generative speech models~\cite{VoiceBox, kim2021conditional, shimizu2024prompttts++} have enabled the synthesis of high-fidelity audio content that closely resembles real-world speech.
Among these, voice cloning techniques~\cite{qin2023openvoice, zhang2022paddlespeech, ruggiero2021voice} can replicate a speaker’s unique vocal characteristics from just a few audio samples, facilitating personalized content generation. When combined with lip synchronization methods~\cite{Wav2Lip, Diff2Lip, PC-AVS} that generate photorealistic video sequences aligned with input speech, these technologies enable the creation of highly realistic audiovisual content. 
Such advancements have broad applications in digital media, virtual avatars, language dubbing, and assistive technologies, where scalable, high-quality audio-visual synchronization is crucial.

However, these techniques also pose significant risks, particularly in the spread of highly realistic misinformation. 
Especially, Synthesized Audiovisual Forgeries (SAVFs)—videos in which both the audio and visuals have been manipulated through voice cloning and lip synchronization—can be exploited to impersonate individuals, manipulate public opinion, and undermine trust in digital media~\cite{cheng2023transface, yang2024synctalklip}.
To mitigate these risks, efforts have focused on detecting and localizing fake visual or audio content in SAVFs. Note that one approach is the localization of tampered regions~\cite{editguard, sanroman2024proactive}, which can help in understanding the attacker's intent and may enable the partial reuse of manipulated content. While localization provides valuable insights, it remains insufficient for assessing the significance of the altered regions or determining the extent of semantic shifts compared to the authentic audio. As a result, it has limitations in fully capturing the attacker's intent and does not allow for the complete reuse of the content.


In this work, we propose a novel task of authentic audio recovery from SAVFs to address the limitations of existing approaches.
Instead of merely detecting or localizing manipulated content, our goal is to reconstruct the authentic audio signal, which directly conveys the semantic content of the message. To achieve this, we introduce a watermarking-based approach that embeds the authentic audio into visual frames before any potential forgery.
This cross-modal approach enables the recovery of authentic audio even when the audio is partially or entirely removed during the forgery process, where direct restoration is particularly challenging.
Beyond audio reconstruction, our method also aids in localizing tampered regions by detecting manipulated content from recovered audio, as manually comparing and identifying altered parts can be highly laborious. 

Through extensive experiments, we show that our approach enables the robust recovery of authentic audio, even when the audio stream is altered or replaced. Furthermore, it precisely localizes tampered regions, providing a proactive defense against audiovisual misinformation. Notably, our approach remains robust even when trained on datasets without human faces or voices, addressing privacy and portrait rights concerns. The core contributions of this paper can be summarized as follows:


\begin{itemize}
\item We introduce the novel task of recovering authentic audio from SAVFs, moving beyond detection and localization to restore authentic audio content.
\item We propose a cross-modal watermarking framework that embeds audio into visual frames, ensuring robust tamper localization and audio recovery even after various manipulations.
\item Experimental results demonstrate that our method enables robust recovery of authentic audio and extends beyond human faces and voices.

\end{itemize}

\begin{figure*}[t]
    \centering
    
    \centering
    \scalebox{1.0}{
        \includegraphics[width=\linewidth]{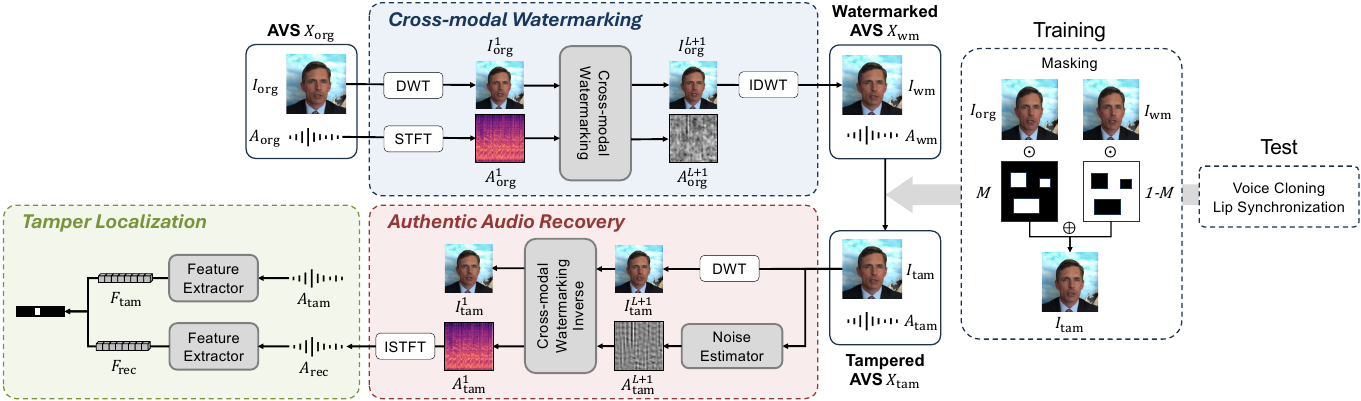}
    }
    \vspace{-0.6cm}
    \caption{
        \textbf{Overall architecture of our model.} The framework comprises three main processes: cross-modal watermarking (CMW), authentic audio recovery, and tamper localization. In the CMW process, CMW embed the authentic audio within a visual frame. For authentic audio recovery, noise estimators predict the transformed audio output from watermarked visual frame, enabling the inverse CMW to recover the authentic audio embedded in the visual frame. Finally, in tamper localization, we compute feature maps for both the recovered and tampered audio to generate a score that identifies the tampered regions.
    }
    \vspace{-0.5cm}
    \label{fig:overview}
\end{figure*}

\section{Related Works}
\label{sec:related}

\noindent\textbf{Voice Cloned Audio Localization}   \ \
Recent advances in voice cloning have intensified the challenge of localizing manipulated audio segments. Approaches like BAM~\cite{BAM} and CFPRF~\cite{wu2024coarse} utilize boundary-aware attention and coarse-to-fine refinement to detect tampering; however, their reliance on specific training manipulations limits robustness against novel attacks. Proactive watermarking methods such as Wavmark~\cite{chen2023wavmark} and Audioseal~\cite{sanroman2024proactive} verify embedded watermarks to identify altered regions, yet they fall short of recovering the authentic audio.


\noindent\textbf{INN-based Steganography and Watermarking}  \ \
Steganography and watermarking embed data within cover content for secure or traceable transfer. Traditional techniques~\cite{LSB, DWT} are constrained by capacity and invisibility issues, leading to the emergence of deep learning methods like~\cite{Steganogan}. Invertible Neural Networks (INNs)~\cite{Nice} offer precise embedding and extraction, as evidenced by HiNet~\cite{hinet}, 
LF-VSN~\cite{LF-VSN} and ThinImg~\cite{thinimg}—the latter hiding audio within images via mel-spectrograms. 
\vspace{-0.1cm}
\section{Method}
\vspace{-0.05cm}
\label{sec:methods}



\subsection{Overview}
\vspace{-0.05cm}
In this work, 
we particularly focus on SAVFs utilized to manipulate the original message of the speaker by modifying real speech videos through voice cloning and lip synchronization.
Our task prioritizes Authentic Audio Recovery (AAR) and Tamper Localization in Audio (TLA) because speech conveys core semantic content, serving as the primary medium for delivering factual and persuasive information.
Formally, our goal is twofold: (1) to recover the authentic audio signal from tampered audiovisual stream and (2) to identify time intervals $\{(t^i_\mathrm{start}, t^i_\mathrm{end})\}_{i=1}^N$ that correspond to tampered audio regions where each pair $(t^i_\mathrm{start}, t^i_\mathrm{end})$ represents a timestamp range indicating a segment of containing tampered content.

To address the tasks of AAR and TLA, we propose a watermarking-based approach, as illustrated in \cref{fig:overview}. 
For notational simplicity, we assume that video $V$ consists of a single visual frame $I$ and the audio segment $A$ that occurs during its display interval.
It is important to note that our technique, formulated under this simplified setting, can be seamlessly extended to Audiovisual Stream (AVS)  of arbitrary length by segmenting them into corresponding visual frames and audio segments.
Specifically, given an original AVS $X_\text{org} = (I_\text{org}, A_\text{org})$ as input, where $I_\text{org}$ is visual frame and $A_\text{org}$ is its corresponding authentic audio, $A_\text{org}$ is imperceptibly embedded into the visual frame $I_\text{org}$, producing a watermarked AVS $X_\text{wm} = (I_\text{wm}, A_\text{wm})$. In this process, $I_\text{wm}$ contains information of $A_\text{org}$, while the audio remains unchanged, indicating that $A_\text{wm}$ is identical to $A_\text{org}$. This watermarked AVS $X_\text{wm}$ can be tampered with methods like lip synchronization and voice cloning, resulting in a tampered AVS $X_\text{tam} = (I_\text{tam}, A_\text{tam})$. For AAR and TLA, we estimate $A_\text{rec}$, a recovered version of $A_\text{org}$, from $I_\text{tam}$. By comparing $A_\text{rec}$ and $A_\text{tam}$, we can localize tampered region. The detailed method for both processes will be elaborated in the following sections.

\subsection{Audio Embedding with Cross-Modal Watermarking} \ \ In our proposed approach, we embed the authentic audio signal $A_\text{org}$ into the visual frame $I_\text{org}$ using Cross-Modal Watermarking (CMW).
To embed $A_\text{org}$ as a watermark into $I_\text{org}$, we adopt trainable Invertible Neural Network (INN) blocks~\cite{LF-VSN} with the unique property of reversibility, allowing for exact recovery of inputs from outputs.

Formally, given an AVS with a single frame $X_\text{org}=(I_\text{org}, A_\text{org})$, our cross-modal watermarking module constructs a watermarked AVS $X_\text{wm}=(I_\text{wm}, A_\text{wm})$.
An INN block at layer $l$ takes inputs $I_\text{org}^l$ and $A_\text{org}^l$ and produces $I_\text{org}^{l+1}$ and $A_\text{org}^{l+1}$ as:
\begin{align}
    I_\text{org}^{l+1} &= I_\text{org}^{l} + \phi(A_\text{org}^{l}), \\
    A_\text{org}^{l+1} &= A_\text{org}^{l} \odot \mathrm{exp}(\sigma(\rho(I_\text{org}^{l+1}))) + \eta(I_\text{org}^{l+1}),
\end{align}
where $\phi, \rho$ and $\eta$ are neural networks, $\sigma$ is a sigmoid activation, and $\odot$ denotes element-wise multiplication.
Note that the inversion operation does not require the direct inverses of $\phi, \rho$ and $\eta$. Instead, it relies on the ability to recover the original inputs from the outputs through the following inversion equations:
\begin{align}
    A_\text{org}^{l} &= (A_\text{org}^{l+1} - \eta(I_\text{org}^{l+1})) \odot \mathrm{exp}(-\sigma(\rho(I_\text{org}^{l+1}))), \\
    I_\text{org}^{l} &=I_\text{org}^{l+1} - \phi(A_\text{org}^{l}).
\end{align}
The only constraints for this inversion are that the output shapes of $\rho$ and $\eta$ must match $A_\text{org}^{l}$, and the output shape of $\phi$ must match $I_\text{org}^{l}$. These conditions ensure that the original inputs can be exactly recovered from the transformed outputs, which is essential for the watermarking operation.

In our implementation, we construct $I_\text{org}^1$ by applying the Discrete Wavelet Transform (DWT) in $I_\text{org} \in \mathbb{R}^{H\times W\times 3}$, where $H$ and $W$ denote the height and width of visual frame. To address the shape discrepancy between $I_\text{org}^1 \in \mathbb{R}^{\frac{H}{4}\times \frac{W}{4}\times 48}$ and $A_\text{org} \in \mathbb{R}^{S}$, where $S$ is length of corresponding audio segment, we transform $A_\text{org}$ into spectrogram by applying the Short-Time Fourier Transform (STFT) and reshape it into $\mathbb{R}^{\frac{H}{4}\times \frac{W}{4}\times 1}$, resulting in $A_\text{org}^1$. Then, $I_\text{org}^1$ and $A_\text{org}^1$ are fed into a stack of L INN blocks.
This process generates outputs $I_\text{org}^{L+1}$ and $A_\text{org}^{L+1}$, with $I_\text{wm}$ obtained by applying Inverse DWT (IDWT) to $I_\text{org}^{L+1}$. 

For $\phi, \rho$ and $\eta$, we 
follow the design principles in~\cite{densenet} using five $3\times3$ convolutional layers with Leaky ReLU activation, where the output channel sizes are adjusted accordingly to match those of $I_\text{org}^l$ and the reshaped $A_\text{org}^l$.
\label{sec:Recovery}
\subsection{Authentic Audio Recovery and Tamper Localization}

\noindent\textbf{Authentic Audio Recovery} \ \ Thanks to the invertible property of INN, we can perfectly recover the original inputs $I_\text{org}$ and $A_\text{org}$ from their outputs $I_\text{org}^{L+1}$ and $A_\text{org}^{L+1}$. 
While $I_\text{org}^{L+1}$ is accessible from the watermarked image $I_\text{wm}$, $A_\text{org}^{L+1}$ cannot be directly accessed from the watermarked AVS $X_\text{wm}$, as it is discarded during the CMW process.
To overcome this, we predict $A_\text{org}^{L+1}$ from $I_\text{wm}$ using a noise estimator inspired by~\cite{LF-VSN}.
This estimation process enables the recovery of the embedded original audio $A_\text{org}$ solely from the watermarked frame $I_\text{wm}$ without requiring access to the authentic audio.
In practice, the watermarked frame $I_\text{wm}$ may also be altered within the SAVFs due to lip synchronization forgery, introducing an additional challenge. 
To address this, we learn a robust model capable of handling such modifications, which will be discussed later.

\label{sec:localization}
\noindent \textbf{Tamper Localization} \ \ After AAR, we can localize the tampered region within the SAVFs by comparing the recovered and tampered audio. 
However, a na\"ive direct comparison of raw audio signals is highly susceptible to noise and recovery errors, making it unreliable for precise tampering localization.
To address this, we use a Semantic Feature Extractor (SFE) based on~\cite{hershey2017cnn} to project the audio streams $A$ into a semantic feature space $F  \in \mathbb{R}^{T \times C}$. We then compute their similarity using the inner product for more robust and reliable TLA.
Formally, given audio feature map $F_\text{tam}=\{\mathbf{f}_\text{tam}^t\}$ and $F_\text{rec}=\{\mathbf{f}_\text{rec}^t\}$ extracted from the $A_\text{tam}$ and $A_\text{rec}$, we compute a cosine similarity score $s^t_\text{tam}=(\mathbf{f}_\text{tam}^t){^\intercal}( \mathbf{f}_\text{rec}^t)$ at timestep $t$, where $\mathbf{f}_\text{tam}^t\in\mathbb{R}^{C}$ and $\mathbf{f}_\text{rec}^t\in\mathbb{R}^{C}$ are temporally aligned feature vectors.
This feature-level comparison enhances resilience to recovery errors and minor perturbations that do not alter the underlying semantics.
We conducted experiments to evaluate the effectiveness of feature-level comparison. While a na\"ive direct comparison of raw audio signals yields an Average Precision (AP) of 87.17, our approach significantly improves it to 98.28, demonstrating superior robustness in tamper localization.

\subsection{Training}
\label{sec:training}
We train the entire network end-to-end in an unsupervised manner, without requiring localization annotations for tampering attacks.
Our total loss function is a weighted sum of four components described below: watermarking loss, visual reconstruction loss, audio reconstruction loss, and feature contrastive loss, formulated as 
$\mathcal{L} = \lambda_\text{WL} \mathcal{L}_\text{WL} + \lambda_\text{VRL} \mathcal{L}_\text{VRL} + \lambda_\text{ARL} \mathcal{L}_\text{ARL} + \lambda_\text{SFCL} \mathcal{L}_\text{SFCL}$ where $\lambda$s are coefficients for each loss term.

\noindent \textbf{Watermarking Loss} \ \ 
To ensure the watermarked visual frame $I_\text{wm}$ closely resembles the original visual frame $I_\text{org}$, we apply an $L_2$ loss, $\mathcal{L}_\text{WL} = || I_\text{wm} - I_\text{org} ||_2^2$.

\noindent \textbf{Reconstruction Losses} \ \ 
To ensure the original frame $I_\text{org}$ and audio stream $A_\text{org}$ are accurately recovered with $A_\text{org}^{L+1}$ missing, we introduce two loss terms $\mathcal{L}_\text{VRL} = || I_\text{org} - I_\text{rec} ||_2^2$ and $\mathcal{L}_\text{ARL} = || A_\text{org} - A_\text{rec} ||_2^2$. 
These terms are critical to train the noise estimator for $A_\text{org}^{L+1}$ introduced in \cref{sec:Recovery}.



\noindent \textbf{Semantic Feature Contrastive Loss} \ \ 
To ensure robust tamper localization, we compare the tampered and recovered audio streams in a semantic feature space. 
Specifically, we enforce proximity between temporally aligned features 
$\mathbf{f}_\text{org}^t$ and $\mathbf{f}_\text{rec}^t$ using a contrastive loss~\cite{InfoNCE} as follows:
\begin{equation}
    \mathcal{L}_{\text{SFCL}} = \sum_{t}\mathcal{L}_{\text{NCE,}t} =-\sum_{t}\log{\frac{\exp{(\mathbf{f}_\text{org}^t \cdot \mathbf{f}}_\text{rec}^t / \tau)}{\sum_{l=1}^{T} \exp{(\mathbf{f}_\text{org}^t \cdot \mathbf{f}_\text{rec}^l / \tau)}}}\label{eqn:cont} \nonumber
\end{equation}
where $\tau$ is a temperature, and $T$ is the number of features in the feature map. 

\noindent \textbf{Masking Strategy} \ \ Lip synchronization forgery alters facial regions, removing parts of embedded watermarks and complicating AAR. 
To enhance robustness against such forgery, we introduce masking strategies during training that partially remove embedded watermarks in $I_\text{wm}$.
As shown in \cref{fig:overview}, given an original frame $I_\text{org}$ and its watermarked counterpart $I_\text{wm}$, we apply a binary mask $M \in \mathbb{R}^{H \times W}$ to selectively replace watermarked regions using $M \odot I_\text{org} + (1 - M) \odot I_\text{wm}$, where $\odot$ denotes element-wise multiplication. 
This simulates watermark removal, helping the model learn to recover audio even when portions are erased.
Specifically, we employ two masking strategies: Random Mask Generation, which applies one to three randomly shaped geometric masks with side lengths sampled between 20 and 150 pixels, and Facial Mask Generation, which uses a facial detection model~\cite{serengil2020lightface} to identify and alter facial regions. In inference time, we utilize lip synchronization models without applying masking strategies.

\begin{table*}[t]
    \centering
    \setlength{\dashlinedash}{2pt}
    \setlength{\dashlinegap}{1pt}
    \caption{\textbf{Comparison of Different Audio Tamper Localization Methods on the HDTF Dataset.} Tampering simulation uses two methods: \textbf{AS}-inserting a different audio segment into the parts of original audio and \textbf{VS}-modifying parts of the audio with voice generated by OpenVoice~\cite{qin2023openvoice}. Localization metrics include IoU, AP, and AUC, while SNR and PESQ measure recovered audio quality, respectively. SSIM and PSNR measure video quality. \textbf{WM} and \textbf{CM} refer to watermarking and cross-modal techniques.}
    \vspace{-0.2cm}
    \scalebox{0.85}{
        \begin{tabular}{
            l| 
            c: 
            c 
            |c 
            c 
            |c 
            c 
            c| 
            c 
            c 
            c| 
            c 
            c| 
            c 
            c 
        }
            \hline
            \hline
            & & & \multicolumn{2}{c|}{Audio Recovery} & \multicolumn{3}{c|}{Tamper Localization (AS)} & \multicolumn{3}{c|}{Tamper Localization (VC)} &  \multicolumn{2}{c|}{Audio Quality} & \multicolumn{2}{c}{Visual Quality}  \\
            \cline{4-15}
            Name & WM & CM & SNR\(\uparrow\) & PESQ\(\uparrow\) & IoU\(\uparrow\) & AP\(\uparrow\) & AUC\(\uparrow\)  & IoU\(\uparrow\) & AP\(\uparrow\) & AUC\(\uparrow\)  & SNR\(\uparrow\) & PESQ\(\uparrow\) & PSNR\(\uparrow\) & SSIM\(\uparrow\) \\
            \hline
            CFPRF~\cite{wu2024coarse} & \ding{55} & \ding{55} & N/A & N/A & 35.12  & 39.21  & 49.12 & 31.31  & 39.77 & 47.58 &  $\infty$ & 4.5 &$\infty$ & 1.0\\
            BAM~\cite{BAM}  & \ding{55} & \ding{55}  & N/A & N/A & 20.43 & 48.44 &  53.82   &  27.24 & 48.29 & 52.42 & $\infty$ & 4.5 & $\infty$ & 1.0 \\
            Wavmark~\cite{chen2023wavmark} & \ding{51} & \ding{55} & N/A & N/A & 40.22 & 40.22 & 50.22  & 40.00 & 40.00 & 50.00 & 36.9 & 4.23 & $\infty$ & 1.0\\
            Audioseal~\cite{sanroman2024proactive} & \ding{51} & \ding{55} & N/A & N/A & 93.68 & 98.23 & 99.02  & 91.78 & 97.43 & 98.69  & 26.5 &  4.39 & $\infty$ & 1.0 \\
            \cdashline{1-15}[2pt/1pt]
            \noalign{\vskip 2pt}
             Ours & \ding{51} & \ding{51} & 17.82 & 3.18 & 97.02& 99.89 & 99.95  & 95.40  & 98.28 & 98.83 & $\infty$ & 4.5 & 41.53 & 0.98 \\
            \hline
            \hline
        \end{tabular}
    
    }

    \vspace{-0.5cm}
    \label{tab:audio_localization}

\end{table*}

\section{Experiments}

\subsection{Experimental Setup}

\noindent \textbf{Dataset} \ \ 
We use the HDTF dataset~\cite{hdtf}, which consists of 410 talking face videos with synchronized speech, totaling 16 hours of audiovisual data. 
As one of the primary benchmarks for lip-synchronization~\cite{hdtf, afouras2018lrs3}, HDTF provides high-quality, diverse speaker recordings, making it well-suited for evaluating the effectiveness of our approach in AAR and TLA.
A random subset of 98 videos is used for training, with the remaining 312 reserved for evaluation.
A random 5-second segments from each of these samples are pre-selected for fair evaluations.
All videos are processed at 25 fps with audio sampled at 16 kHz.

\begin{table}[t]
    \centering
    \caption{\textbf{Impact of Masking Strategy} “No Mask” refers to training without a mask. Our masking strategies outperform “No Mask”, showing their effectiveness.}
    \vspace{-0.2cm}
    \scalebox{0.75}{
        \begin{tabular}{
            l| 
            c 
            c| 
            c 
            c| 
            c 
            c 
        }
            \hline
            \hline
             Masking & \multicolumn{2}{c|}{Audio Recovery} & \multicolumn{2}{c|}{Tamper Localization} &  \multicolumn{2}{c}{Visual Quality} \\
            \cline{2-7}
            Strategy & SNR\(\uparrow\) & PESQ\(\uparrow\) & IoU\(\uparrow\) & AUC\(\uparrow\) & PSNR\(\uparrow\) & SSIM\(\uparrow\) \\
            \hline
            No Mask & 4.63 & 1.53 & 60.01 & 86.10 &  \textbf{42.56} & \textbf{0.99} \\
            Facial Mask  &  \textbf{18.17} & \textbf{2.73} & \textbf{95.19}  &\textbf{99.26} & 41.53 & 0.98 \\
            Random Mask &  17.41 & 2.36 &  92.29 & 98.88 & 40.64  & 0.98 \\
            \hline
            \hline
        \end{tabular}
        
    }

    \vspace{-0.25cm}
    \label{tab:mask_effect}
\end{table}

\noindent \textbf{Evaluation Metrics} \ \ 
For AAR, we adopt Signal-to-Noise Ratio (SNR), Perceptual Evaluation of Speech Quality (PESQ) from~\cite{sanroman2024proactive, chen2023wavmark}, which are common metrics, measuring the audio fidelity.
For TLA, we employ Intersection over Union (IoU), Average Precision (AP), and Area Under the Curve (AUC) following~\cite{editguard, zhang2024v2a}.
We additionaly measure the quality of the watermarked contents using Peak Signal-to-Noise Ratio (PSNR) and Structural Similarity Index Measure (SSIM) from~\cite{editguard}.

\noindent \textbf{Implementation Details} \ \ 
We use six invertible blocks ($L = 6$) and optimize the weights for 10K iterations using Adam optimizer with a learning rate of $1 \times 10^{-4}$, $\beta_1=0.9$ and $\beta_2=0.5$. 
For all experiments, we set $\lambda_{\mathrm{WL}} = 10$, $\lambda_{\mathrm{ARL}} = 10$, $\lambda_{\mathrm{VRL}} = 0.1$,  $\lambda_{\mathrm{SFCL}} = 1$ and $\tau=0.07$ while setting the window size and hop length to 510 and 128 for STFT. The channel of the audio feature map is 32 and we employ a queue that stores 65,536 features from previous iterations.

\subsection{Results}

\begin{figure}[t]
    \centering
    
    \centering
    \scalebox{1}{
        \includegraphics[width=\linewidth]{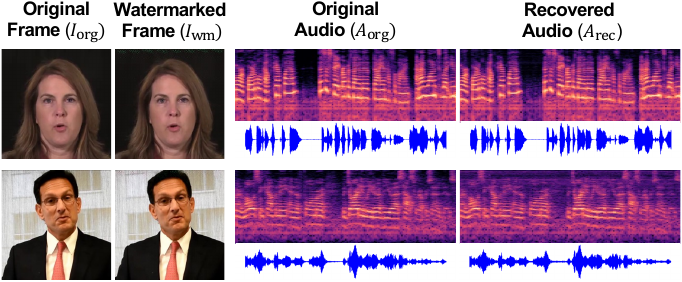}
    }
    \vspace{-0.6cm}
    \caption{
        \textbf{Qualitative Examples} The watermarked frames and recovered audio closely resemble the original AVS, ensuring imperceptible embedding and authentic audio recovery.
    }
    \label{fig:qualitative}
    \vspace{-0.5cm}
\end{figure}

\noindent \textbf{Tamper Localization in Audio} \ \ 
\cref{tab:audio_localization} compares our model with SOTA audio tamper localization methods, including passive approaches, BAM~\cite{BAM} and CFPRF~\cite{wu2024coarse} and proactive watermarking-based (WM) methods, AudioSeal~\cite{sanroman2024proactive} and Wavmark~\cite{chen2023wavmark}.
We evaluate two tampering scenarios: Audio Swapping (AS), where 10\% to 30\% of audio is replaced with another segment from the same speaker, and Voice Cloning (VC), where a audio segment is substituted with a synthetic voice generated by a cloning model~\cite{qin2023openvoice}.
In both cases, visual frames are also manipulated using MuseTalk~\cite{musetalk}, posing an additional challenge for our model.
Unlike audio-only baselines that ignore visual inputs, our model embeds audio within the visual modality, enabling accurate recovery and tamper localization.
As a result, our cross-modal watermarking model consistently outperforms the best SOTA method, Audioseal, in both settings and surpasses other baselines by a large margin.
Unlike Wavmark and Audioseal, which degrade audio quality by embedding watermarks directly in the audio stream, our approach preserves audio integrity by embedding the watermark into the visual input while maintaining high visual quality. \cref{fig:qualitative} shows that this process has minimal impact on visual quality.
\begin{table}[t]
    \centering
     \caption{\textbf{Comparison of Different Lip Synchronization Methods.} Lip synchronization is simulated with Wav2Lip~\cite{Wav2Lip}, Diff2Lip~\cite{Diff2Lip}, and MuseTalk~\cite{musetalk}. The watermark is effectively extracted after lip synchronization.}
     \vspace{-0.25cm}
     
\scalebox{0.85}{
        \begin{tabular}{
            l| 
            c 
            c| 
            c 
            c 
            c 
        }
            \hline
            \hline
            Lip synchronization & \multicolumn{2}{c|}{Audio Recovery} & \multicolumn{3}{c}{Tamper Localization} \\
            \cline{2-6}
            Model &  SNR\(\uparrow\) & PESQ\(\uparrow\) & IoU\(\uparrow\) & AP\(\uparrow\) & AUC\(\uparrow\)  \\
            \hline
            None & 28.20 & 3.49 & 97.13 & 99.71 & 99.80 \\
            Wav2Lip~\cite{Wav2Lip} & 16.06 & 2.91 & 92.97 & 95.23 & 98.29 \\
            Diff2Lip~\cite{Diff2Lip} & 18.17 & 2.73 & 95.19 & 99.02 & 99.26  \\
            MuseTalk~\cite{musetalk}  & 17.82 & 3.18 & 95.40 & 98.28 & 98.83 \\
            \hline
            \hline
        \end{tabular}
    }

    \vspace{-0.3cm}
    \label{tab:lip_synchronization}
    
\end{table}
\begin{table}[t]
    \centering
    \caption{\textbf{Domain Generalization to Unseen Domains during Training.
} Comparisons of our method trained on human-associated~\cite{hdtf} and non-human~\cite{xue2019video, FMA} datasets, demonstrating its domain generalization capability.}
    \vspace{-0.25cm}  
    \scalebox{0.65}{
        \begin{tabular}{
            l| 
            c 
            c| 
            c 
            c| 
            c 
            c 
        }
            \hline
            \hline
             Training & \multicolumn{2}{c|}{Audio Recovery} & \multicolumn{2}{c|}{Tamper Localization} &   \multicolumn{2}{c}{Visual Quality} \\
            \cline{2-7}
            Dataset  & SNR\(\uparrow\) & PESQ\(\uparrow\)& IoU\(\uparrow\)  & AUC\(\uparrow\) & PSNR\(\uparrow\) & SSIM\(\uparrow\) \\
            \hline
            HDTF~\cite{hdtf} & 17.41 & 2.36 & 92.29 & 98.88  &  40.64 & 0.98 \\
            Vi90k + FMA~\cite{xue2019video, FMA} & 15.40 & 1.82 & 88.26 & 97.72  & 40.49 & 0.98 \\
            \hline
            \hline
        \end{tabular}
        
    }
    \vspace{-0.7cm}
    
    \label{tab:out_of_domain}
\end{table}


\noindent \textbf{Authentic Audio Recovery} \ \  
\cref{tab:audio_localization} shows that, despite their tamper localization capabilities, All baseline models are unable to perform AAR. In contrast, our model, leveraging a robust cross-modal watermarking technique, successfully recovers the authentic audio, achieving an SNR of 17.82 and a PESQ of 3.18 from SAVFs, demonstrating outstanding AAR performances. This is confirmed by the qualitative examples in~\cref{fig:qualitative}. 

\noindent \textbf{Effects of Masking Strategies} \ \ 
\cref{tab:mask_effect} compares different masking strategies during training, highlighting their impact on AAR and TLA. 
Training without masking fails to recover authentic audio, as watermarks are fragile to lip synchronization forgery, making masking essential for AAR.
Both random and facial masking improve robustness, with facial masking achieving higher performances since its masked region aligns with the actual tampered areas, enabling more effective signal embedding. 
In terms of visual quality, training without masking preserves visual fidelity best, as it allows for compact signal embedding but lacks robustness to attacks. 
Among masking strategies, facial masking maintains better visual quality than random masking, as it introduces less redundancy due to its more predictable regions, while random masking increases uncertainty, leading to more dispersed watermark embedding.



\noindent \textbf{Robustness to Various Lip Synchronization Methods} \ \ 
We evaluate our model's robustness against three lip synchronization methods \cite{Wav2Lip, Diff2Lip, musetalk}. 
As shown in \cref{tab:lip_synchronization}, performance is highest without lip synchronization, preserving the embedded watermark.
While lip synchronization causes some degradation, speech remains intelligible (refer to Supplementary Material), and scores remain relatively high compared to the failed model in \cref{tab:mask_effect}.
Additionally, scores for TLA consistently exceed 90\%.
These results highlight our model’s effectiveness in SAVF scenarios and demonstrate its reliability under various adversarial conditions.





\noindent \textbf{Learning from Non-Human Datasets}
Training networks with human-associated datasets often raises ethical concerns, particularly regarding privacy.
To mitigate these issues, we train our model using non-human datasets that exclude human facial imagery and human voice. 
Specifically, we use the Vimeo-90k dataset~\cite{xue2019video}, combined with the Free Music Archive (FMA) dataset~\cite{FMA}.
As shown in Table~\ref{tab:out_of_domain}, our model trained on this dataset achieves slightly lower but comparable performance to the model trained directly on HDTF in both AAR and TLA. 
These results demonstrate that our approach can be effectively trained on datasets from entirely different domains, addressing privacy concerns without significant performance degradation.


\section{Conclusion}
In this paper, we introduced a novel task of recovering authentic audio from SAVFs, moving beyond mere detection and localization. To achieve this, we propose cross-modal watermarking method not only localizing tampered regions but also recovering authentic audio. Our model demonstrated state-of-the-art localization performance while effectively recovering authentic audio. Notably, our approach remains effective without training on human faces or voices, ensuring privacy compliance. This practical solution combats misinformation and preserves content authenticity, fostering a safer multimedia ecosystem.


\section{Acknowledgements}
This research was supported by IITP grants (IITP\allowbreak-2025-RS-2020-II201819, IITP-2025-RS-2024-00436857, IITP-2025-RS-2024-00398115, IITP-2025-RS-2025-02263754, IITP-2025-RS-\allowbreak2025-02304828),  and the KOCCA grant (RS-2024-00345025) funded by the Korea government (MSIT, MOE and MSCT).

\bibliographystyle{IEEEtran}
\bibliography{mybib}

\end{document}